\documentclass[letterpaper]{article} 
\usepackage{aaai2026}  
\usepackage{times}  
\usepackage{helvet}  
\usepackage{courier}  
\usepackage[hyphens]{url}  
\usepackage{graphicx} 
\usepackage{natbib}  
\usepackage{caption} 
\usepackage{algorithm}
\usepackage{algorithmic}
\usepackage{amsmath}
\usepackage{amssymb}
\usepackage{amsthm}
\newtheorem{definition}{Definition}
\newtheorem{example}{Example}
\usepackage{booktabs}
\usepackage{graphicx}
\usepackage{subcaption}
\usepackage{threeparttable}

\usepackage{pifont}
\usepackage{xcolor}
\usepackage{tcolorbox} %
\usepackage{xspace}
\newcommand{\tool}{{\sc TAPA}\xspace}

\newcommand{\commentout}[1]{}

\usepackage{newfloat}
\usepackage{listings}
\DeclareCaptionStyle{ruled}{labelfont=normalfont,labelsep=colon,strut=off} 
\floatstyle{ruled}
\newfloat{listing}{tb}{lst}{}
\floatname{listing}{Listing}
\title{Tapas Are Free! Training-Free Adaptation of Programmatic Agents via LLM-Guided Program Synthesis in Dynamic Environments}
\author {
    Jinwei Hu\textsuperscript{\rm 1},
    Yi Dong\textsuperscript{\rm 1}\thanks{Corresponding author.},
    Youcheng Sun\textsuperscript{\rm 2},
    Xiaowei Huang\textsuperscript{\rm 1}  
}
\affiliations {
    \textsuperscript{\rm 1}School of Computer Science and Informatics, University of Liverpool, Liverpool, L69 3BX, the UK\\
    \textsuperscript{\rm 2}Department of Computer Science, Mohamed bin Zayed University of Artificial Intelligence, Abu Dhabi, SE45 05, UAE\\
    \{jinwei.hu, yi.dong, xiaowei.huang\}@liverpool.ac.uk, youcheng.sun@mbzuai.ac.ae
}

\usepackage{bibentry}

\begin{document}

\maketitle

\begin{abstract}
Autonomous agents in safety-critical applications must continuously adapt to dynamic conditions without compromising performance and reliability. This work introduces \textbf{TAPA} (\textbf{T}raining-free \textbf{A}daptation of \textbf{P}rogrammatic \textbf{A}gents), a novel framework that positions large language models (LLMs) as intelligent moderators of the symbolic action space. Unlike prior programmatic agents typically generate a monolithic policy program or rely on fixed symbolic action sets, \tool synthesizes and adapts modular programs for individual high-level actions, referred to as logical primitives. By decoupling strategic intent from execution, \tool enables meta-agents to operate over an abstract, interpretable action space while the LLM dynamically generates, composes, and refines symbolic programs tailored to each primitive. Extensive experiments across cybersecurity and swarm intelligence domains validate \tool's effectiveness. In autonomous DDoS defense scenarios, \tool achieves 77.7\% network uptime while maintaining near-perfect detection accuracy in unknown dynamic environments. In swarm intelligence formation control under environmental and adversarial disturbances, \tool consistently preserves consensus at runtime where baseline methods fail. This work promotes a paradigm shift for autonomous system design in evolving environments, from policy adaptation to dynamic action adaptation. 
\end{abstract}

\section{Introduction}
Autonomous agents have become increasingly prevalent across critical domains such as cyber defense \cite{lohn2023autonomous}, swarm intelligence control \cite{duan2025enhancing}, and autonomous driving \cite{kiran2021deep}, where they are tasked with making timely and reliable decisions under complex conditions. 
Recent advancements in symbolic AI and reinforcement learning (RL) have greatly enhanced these agents' ability to interact with their environments, enabling the acquisition of structured and abstract behaviors beyond pure rule-based control \cite{carnevali2024neuro}. The integration of neuro-symbolic approaches has emerged as the predominant paradigm for  safety-critical applications, combining neural learning capabilities with symbolic rule reliability to overcome the black-box limitations of pure RL while maintaining the interpretability and formal guarantees required in critical domains \cite{marton2024sympol}. In such hybrid systems, symbolic modules are responsible for concrete execution by employing symbolic programs or rules as elements of the action space, enabling agents termed \textit{programmatic agents} to operate through interpretable programmatic actions \cite{lyu2019sdrl, verma2019imitation}. Compared to traditional actions, programmatic symbolic action spaces provide superior interpretability, verifiability, and integration of expert knowledge, making them well-suited for real-world applications \cite{lelis2024programmatic}.
\begin{figure}[htbp]
    \centering
    \includegraphics[width=\linewidth]{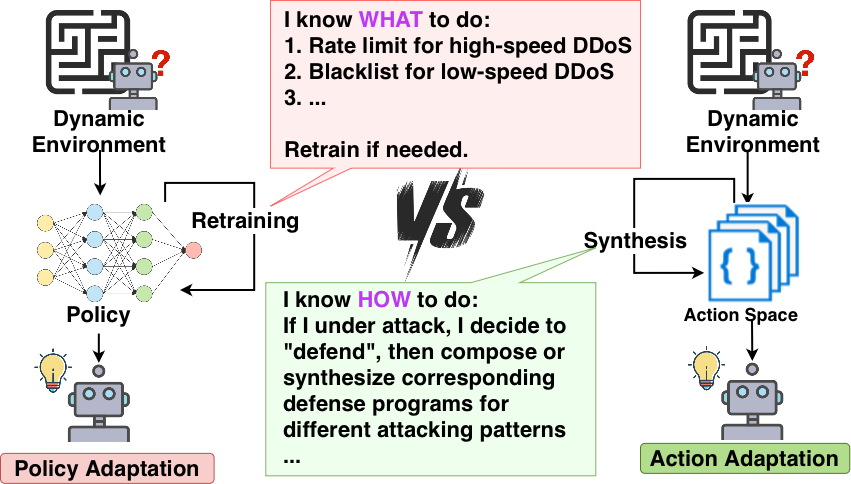} 
    \caption{Policy‑Level Retraining (left) vs. Action‑Level Synthesis and Adaptation (right)}
    \label{fig:strategy_comparison}
\end{figure}

\begin{figure*}[t]
    \centering
    \includegraphics[width=\linewidth]{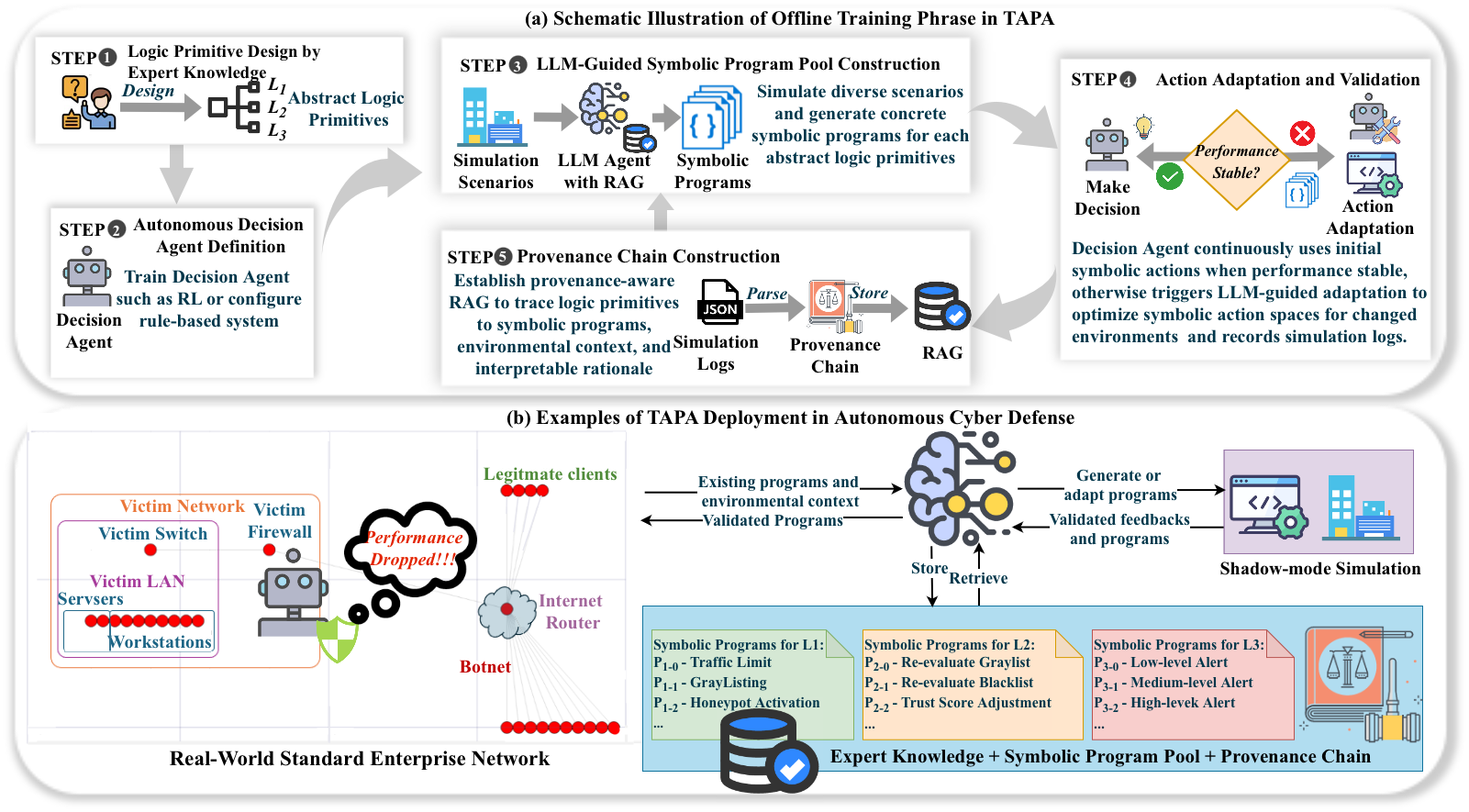} 
    \caption{
    TAPA ({T}raining-free {A}daptation of {P}rogrammatic {A}gents) Framework.
    \textbf{(a) Design-time workflow.} TAPA enables autonomous agents to adapt to evolving environments without retraining through LLM-guided symbolic program synthesis:  
    \emph{\ding{192}~Logic primitive design.} Define high-level symbolic operations based on expert knowledge as interpretable strategic intent.  
    \emph{\ding{193}~Decision agent initialization.} A meta-agent is instantiated to select logical primitives based on environmental conditions.  
    \emph{\ding{194}~LLM-guided program pool construction.} LLM generates diverse symbolic programs across multiple simulated scenarios for each primitive.
    \emph{\ding{195}~Action adaptation and validation.} When performance degrades, LLM synthesizes candidate programs for action adaptation and validated them through shadow simulation before replacement.
    \emph{\ding{196}~Provenance chain construction.} Execution traces and adaptation experiences are stored in a Retrieval-Augmented Generation (RAG) system for future program synthesis.
    \textbf{(b) Deployment-time use case for cyber defense.} The TAPA-enabled agent monitors network performance, detects degradation, and retrieves or synthesizes validated programs as adaptive defensive operation in dynamic environments.
    }
    \label{fig:TAPA-overview}
\end{figure*}
However, these static agent frameworks fundamentally lack \textit{self-adaptive updating capabilities} and struggle to match the continuous changed characteristic of dynamic operational environments \cite{hu2025stopreducingresponsibilityllmpowered}. The rapid evolution of attack vectors, environmental conditions, and task requirements in safety-critical scenarios further exacerbates these limitations, resulting in degraded performance and rigid decision-making that overlooks the crucial need for adaptiveness to rapidly respond to unforeseen changes and urgent threats \cite{bhuyan2024neuro, li2024logicity,hu2025trustorientedadaptiveguardrailslarge}. Therefore, we summarized the existing approaches fall short in three aspects: (i) limited adaptability due to rigid symbolic action spaces; (ii) prohibitive retraining costs requiring extensive strategy overhaul or manual rule updates; and (iii) inadequate safeguards when outdated symbolic actions are applied to changed environments.

Although Large Language Models (LLMs) offer new possibilities with their powerful reasoning and generalization capabilities \cite{hadi2023survey,guo2024large}, current research largely focuses on end-to-end LLM decision agent, overlooking their potential to agents' adaptiveness. Moreover, LLMs struggle in safety-critical, latency-sensitive environments, such as Distributed Denial of Service (DDoS) attack mitigation in cybersecurity scenarios where their inference overhead and the possibility of hallucinated output lead to inferior real-time performance compared to well-trained RL agents \cite{hager2024evaluation, castro2025large}. This reveals a critical gap where existing autonomous systems lack adaptability, and LLMs lack efficiency and reliability. Motivated by these complementary issues, we pursue a new paradigm shown in Figure~\ref{fig:strategy_comparison}: \textit{Can we harnesses LLMs' generative and reasoning capabilities for dynamic program synthesis and action space moderation, enabling rapid adaptation while preserving efficiency and reliability of well-trained policies or expert-defined configurations?}


To address this challenge, we propose the {TAPA} framework, illustrated in Figure~\ref{fig:TAPA-overview}. Overall, the key contributions of this paper are summarized as follows:
\begin{itemize}
\item We introduce LLMs as intelligent action space moderators for autonomous agents, achieving a paradigm shift from costly policy to efficient action adaptation.

\item We propose TAPA, a training-free framework enabling agents to adapt to evolving environments through LLM-guided symbolic program synthesis while enhancing adaptability, interpretability and traceability.

\item We validate TAPA across safety-critical scenarios including cyber defense and swarm control, demonstrating superior performance in DDoS mitigation and formation control under adversarial disturbances.
\end{itemize}

\section{Related Work}
\subsection{Autonomous Decision Agents}
Autonomous intelligent agents for real-time decision-making predominantly rely on traditional planning methods, reinforcement learning, or LLMs, each exhibiting critical limitations in dynamic environments requiring rapid adaptation. Traditional planning approaches such as STRIPS \cite{bylander1994computational, aineto2018learning}, PDDL-based planners \cite{holler2020hddl}, and Hierarchical Task Networks \cite{hogg2009learning, kelly2008offline} require comprehensive domain specifications with predefined action sets, relying heavily on expert knowledge for domain modeling. While theoretically robust, their static nature and labor-intensive requirements significantly constrain practicality when rapid operational adaptations are required \cite{bhuyan2024neuro}.

Neuro-symbolic approaches integrate RL with symbolic AI to enhance capabilities through environmental interaction. These hybrid methods allow RL agents to operate over symbolic action spaces, or guide symbolic planning using learned value functions \cite{yang2018peorl, lyu2019sdrl, shindo2025blendrl}. However, they encounter critical limitations in adaptability under adversarial or highly dynamic conditions, where symbolic priors become obsolete and RL agents struggle to generalize across different operational scales due to rigid logic constraints \cite{hakim2025ansr,11077439}. End-to-end LLM agents leverage natural language understanding to directly produce action sequences, enabling flexible reasoning and context-aware interaction \cite{yao2023react}. Despite their versatility, these agents face critical limitations in safety-critical deployments: inference latency hinders real-time responsiveness \cite{liang2025llm}, and outputs are prone to uncertainty and hallucination, undermining reliability even when guardrails are equipped \cite{pmlr-v235-dong24c,dong2024safeguardinglargelanguagemodels}. In contrast, our TAPA framework positions LLMs as intelligent action space moderators, dynamically adapting action via programs synthesis. This approach addresses the adaptability-efficiency trade-off by enabling real-time symbolic program evolution without costly retraining, allowing agents to safely handle arbitrary scenarios including unexpected emergent situations through symbolic program adaptation.
\subsection{Large Language Models for Code Generation}
Recent advances in LLMs have revolutionized code generation capabilities, with models like GPT \cite{achiam2023gpt}, Claude \cite{claude3}, and Code Llama \cite{roziere2023code} demonstrating impressive performance on standard benchmarks such as HumanEval and MBPP \cite{jiang2024survey}. These developments have led to widespread adoption of LLM-based coding assistants in software development workflows, including GitHub Copilot and various IDE integrations \cite{el2024using}. However, code generation in safety-critical domains presents unique challenges that differentiate it from general-purpose programming tasks \cite{soroush2024large}. While RAG-based approaches \cite{parvez-etal-2021-retrieval-augmented,zhou2023docprompting} 
have enhanced code generation with external knowledge, the scarcity of high-quality, domain-specific code samples in safety-critical domains means RAG can only offer limited descriptive guidance, resulting in practically ineffective code generation. Recent work has explored various methods to address domain-specific code generation challenges, including specialized training datasets and domain-specific fine-tuning \cite{guo2024redcode,zan2023large,hou2025enhancing}. However, these approaches primarily focus on improving syntactic and functional correctness rather than ensuring domain-specific operational effectiveness in dynamic environments where code must adapt to rapidly evolving scenarios and threats. To enhance the practicality of generated programs, our framework constructs symbolic program pools through domain-specific multi-scenario simulation and employs provenance chains to guide LLM moderators in making informed decisions for symbolic program selection, composition, adaptation, and even new generation during runtime based on validated execution experiences.

\section{Problem Formulation}

Consider an autonomous agent operating in an environment $E$ characterized by state space $\mathcal{S}$, action space $\mathcal{A} = \{P_1, P_2, ..., P_m\}$ where $m$ denotes the number of available symbolic programs, and reward function $R(s,p)$. The agent maintains policy $\pi: \mathcal{S} \rightarrow \mathcal{A}$ and aims to maximize cumulative rewards:
\begin{equation} \label{eq: reward objective}
   \max_{\pi} \mathbb{E}_{\tau \sim (E, \pi)} \left[ \sum_{t=0}^{T} \gamma^t R(s_t, a_t) \right]
\end{equation}
where $\tau$ represents a trajectory, $T$ is the time horizon, and $\gamma \in [0,1]$ is the discount factor. However, real-world deployment environments for autonomous decision agents are typically dynamic and evolve rapidly, especially in safety-critical scenarios. When the environment evolves across multiple versions $\mathcal{E} = \{E_1, E_2, ..., E_V\}$, where $V$ denotes the number of environment variants, a fixed action space and its corresponding trained policies may become insufficient, thereby limiting the agent's generalizability. Environmental evolution involves multiple dimensions such as scale, complexity, adversarial patterns, and operational constraints, causing significant performance degradation $R_{v+1} \ll R_v$, where $v$ is the index of environmental version. To maintain agent stability and address the aforementioned challenges posed by environmental evolution, we reframe autonomous agent training or configuration by abstracting the concrete action space $\mathcal{A}$ into a set of logical primitives $\mathcal{L} = \{L_1, L_2, ..., L_N\}$, where $N$ denotes the number of available logical actions.

\begin{definition}[Logical Primitive] \label{de: logical primitives}
A logical primitive $L_i \in \mathcal{L}, i \leq N$ represents a high-level strategic intent that specifies \emph{what} the agent should accomplish in a given context, without prescribing \emph{how} to implement. These primitives abstract over concrete actions and symbolic program implementations, enabling decision-making to remain generalizable and interpretable across evolving environments.
\end{definition}

Abstract policies over $\mathcal{L}$ enhance generalizability by decoupling strategic intent from low-level execution, allowing the symbolic layer to adapt independently to environmental changes while preserving policy stability. Given this abstraction, the challenge becomes finding optimal symbolic program mappings for each logical primitive that enable performance stability in evolved environments. For an effective abstract policy $\pi^*: \mathcal{S} \rightarrow \mathcal{L}$ and evolved environment $E_{v+1}$, we seek the optimal mapping:

\begin{equation} \label{eq:symbolic program moderation}
\mathcal{M}^* = \arg\max_{\mathcal{M}} \mathbb{E}[R_{v+1} | \pi^*, E_{v+1}, \mathcal{M}(\mathcal{L})]
\end{equation}
where $\mathcal{M}: \mathcal{L} \rightarrow \mathcal{A}_{mod}$ maps each logical primitive to moderated symbolic programs, where $\mathcal{A}_{mod}$ represents the space of all possible program compositions. Formally, for each logical primitive $L_i$:

\begin{equation}
\mathcal{M}(L_i) = \bigcup_{j} \left( P_j \odot \text{op}_j \right)
\end{equation}
where $P_j \in \mathcal{A}$ are selected symbolic programs, $\odot$ represents composition operations that apply $\text{op}_j$ to programs, and $\text{op}_j \in \{\land, \lor, +, -, \delta\}$ denotes logical operations including conjunction, disjunction, addition, deletion, and modification applied to program $P_j$. This formulation shifts adaptation from expensive policy retraining to efficient symbolic action space moderation, ensuring adaptive decision-making in dynamic scenarios.

\section{TAPA Framework}


In the following, we describe each step of the \tool framework shown in Figure~\ref{fig:TAPA-overview} in details.

\subsubsection{\ding{192} Logic Primitive Design by Expert Knowledge}
We begin by defining a set of domain-specific logical primitives $\mathcal{L}$ according to Definition~\ref{de: logical primitives} based on scenario and task requirements. Each primitive represents a strategic intent derived from domain expertise and serves as interpretable abstractions over concrete actions, enabling more stable and generalizable strategy across varying environments.
\begin{example} [Logical Primitives in Cyber Defense] \label{ex: Logical Primitives}
In cybersecurity scenarios, experts typically define logical primitives as $\mathcal{L} = \{\text{Observe, Defend, Validate, Alert}\}$, where \textit{Observe} monitors network traffic patterns, \textit{Defend} applies countermeasures like filtering and rate limiting, \textit{Validate} verifies threat authenticity, and \textit{Alert} triggers human verification.
\end{example}

\subsubsection{\ding{193} Decision Agent Initialization}
Based on this abstraction, we formally define a meta-agent that operates over $\mathcal{L}$ rather than directly over the raw action space as follows:
\begin{definition}[Meta-Agent and Policy]
A meta-agent is an autonomous agent that operates over abstract logical primitives rather than concrete actions, learning or being configured with a meta policy $\pi_{\text{meta}}: S \rightarrow \mathcal{L}$ based on Eq.~\eqref{eq: reward objective}.
\end{definition}
Under this definition, the meta-policy can be optimized in a simplified environment $E_{\text{simple}}$ by replacing the original action space $\mathcal{A}$ with logical primitives $\mathcal{L}$, thereby enhancing generalizability across evolving environments.

\begin{example} [Meta-Agent in Cyber Defense] \label{ex: meta agent initilization}
Continuing from Example~\ref{ex: Logical Primitives}, a programmatic meta-agent learns to select appropriate logical primitives based on network states. For instance, when detecting abnormal traffic patterns, the meta-policy $\pi_{\text{meta}}(s_t) = \text{Defend}$, triggering the defensive intent rather than directly specifying which symbolic programs or rules to apply.
\end{example}

\subsubsection{\ding{194} LLM-Guided Symbolic Program Pool Construction} 
We firstly construct diverse scenarios across multiple simulated environments to capture various operational conditions and challenges. For each logical primitive $L_i$ and simulated environment $E_v$ from the set of environmental versions $\mathcal{E}$, we employ an LLM augmented with domain-specific expert knowledge to generate candidate symbolic programs. The generation process is formalized as: 
\begin{equation} 
   \mathcal{P}_{i,v} = \text{LLM}_{\text{gen}}(L_i, \xi(E_v), \mathcal{V}_{\text{RAG}}) 
\end{equation}
where $\mathcal{P}_{i,v}$ represents the candidate program set for primitive $L_i$ in environment $E_v$, $\xi(E_v)$ denotes environmental context features, and $\mathcal{V}_{\text{RAG}}$ contains expert knowledge and experience from past simulations. The LLM agent leverages this knowledge to ensure generated programs are domain-compliant and contextually appropriate.

\begin{example} [Program Pool in Cyber Defense] \label{ex: program pool}
Continuing from Example \ref{ex: meta agent initilization}, for the logical primitive $L_1$ across multiple DDoS scenarios, the LLM agent generates a series of concrete programmatic programs such as $\{P_{1,1}, P_{1,2}, P_{1,3}, ...\}$ to construct the program pool. For instance, $P_{1,1}$ implements rate limiting, $P_{1,2}$ implements traffic filtering, and $P_{1,3}$ implements blacklisting and rate limiting.
\end{example}

\subsubsection{\ding{195} Action Adaptation and Validation} 
We conduct multi-scenario simulations to validate the symbolic programs in the constructed program pool. During this process, when performance degradation is detected typically through reward $R(s_t, L_t)$ below threshold $\xi$, the LLM agent leverages its reasoning capabilities to invoke the RAG system and determine optimal symbolic program compositions following Eq.~\eqref{eq:symbolic program moderation} to perform action adaptation by replacing the programs equipped for corresponding logical primitives. This process involves either composing existing programs using logical operations when similar cases exist in the RAG system, or synthesizing new programs guided by RAG-stored experiences when no comparable scenarios are found.

All candidate programs must undergo validation through shadow simulation \cite{kuchta2018shadow, microsoft2023shadow}, a standard practice in industrial deployments where new algorithms run in parallel with production systems without affecting live operations. Within this shadow simulation environment, when significant environmental shifts are detected, the framework can also generate several alternative meta-policies $\{\pi_{\text{meta}}^{(1)}, \pi_{\text{meta}}^{(2)}, ..., \pi_{\text{meta}}^{(k)}\}$ where $k$ represents the number of viable strategic alternatives, as backup options to maintain policy stability and ensure uninterrupted decision-making in safety and time-critical scenarios. This holistic design guarantees safety and performance before deployment while enabling quick strategic switches based on real-time feedback.

\begin{example}[Action Adaptation in Cyber Defense] 
Continuing from Example \ref{ex: program pool}, when performance drops below threshold during DDoS scenarios, action adaptation is triggered. The LLM agent analyzes environmental and performance context (traffic volume: 2.3K packets/sec, network uptime: 22\%) and retrieves similar cases from the RAG system to compose multiple potential programs $\{P_{adapt}^{(1)}, P_{adapt}^{(2)}, ...\}$. After shadow simulation validation shows $P_{adapt}^{(1)} = P_{1,1} \land P_{1,3}$ for $L_1$ can recover performance to 72\% (e.g., acceptable threshold), this adapted program replaces the original to serve as the new execution action for $L_1$ while continuously searching for better-performing actions or backup strategy combinations.
\end{example}

\subsubsection{\ding{196} Provenance Chain Construction}
\begin{figure}[htbp]
    \centering
    \includegraphics[width=1\linewidth]{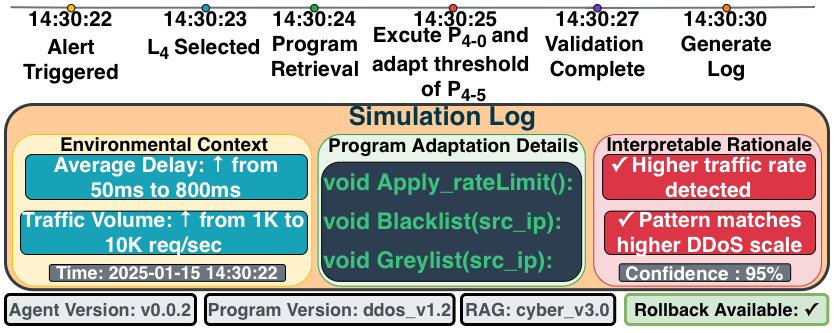}
    \caption{Provenance chain example for DDoS attack.}
    \label{fig:Provenance chain example}
\end{figure}

Each simulation generates a detailed provenance chain (PC) stored in the RAG system as experience for future program synthesis. We record comprehensive execution traces based on simulation logs, including invoked logical primitives, environmental context, original and modified programs, performance profiles, and adaptation rationales as shown in Figure~\ref{fig:Provenance chain example} and Appendix
B. The PC and constructed program pool together enable full traceability, facilitate system debugging, and enhance LLM-driven program generation by leveraging prior adaptation experiences, \textit{addressing the scarcity of domain-specific program samples and experience}. This iterative optimization process across diverse scenarios progressively enriches the RAG system, establishing a comprehensive knowledge base for runtime adaptation.

\section{Experiments}
To evaluate the effectiveness of TAPA across safety-critical domains, we conduct experiments on two representative cases:  
(1) \textit{Autonomous cyber defense}, where agents must rapidly adapt to evolving DDoS attack patterns; and  
(2) \textit{Swarm formation control}, where UAVs must maintain consensus under environmental disturbances and adversarial interference.  
These domains exemplify settings in which static symbolic systems struggle to adapt effectively~\cite{dehghantanha2023autonomous,stodola2025dynamic}. Our results demonstrate that \tool enables training-free action adaptation while preserving interpretability and operational efficiency. 
Implementation details for both domains are provided in Appendix
A. In addition, we perform ablation studies to validate the design rationale behind key components of the framework.

\subsection{Autonomous Cyber Defense}
This cybersecurity defense use case is designed to evaluate the effectiveness of action adaptation under evolving DDoS attack patterns and changing network conditions, and to assess the agent’s ability to maintain high detection accuracy, low false positives, and sustained network uptime (without policy retraining, which typically needs hours to adapt to unknown environments).

\paragraph{Experimental Setup}
We conduct our experiments using the NS3AI simulation platform~\cite{yin2020ns3}, which combines high-fidelity network simulation~\cite{henderson2008network} with Python-based AI frameworks via shared memory linking, enabling real-time simulation and protocol-level traffic modeling. Our network topology mirrors standard enterprise architectures used in cyber defense testbeds (e.g., CIC-IDS2017, CIC-DDoS2019)~\cite{sharafaldin2018toward}, comprising a victim LAN with router/firewall, legitimate TCP clients, and a botnet launching high-rate attacks. To evaluate defense performance, we design five progressively complex scenarios (E2–E5) varying in attack sophistication, network size, and operational constraints, as summarized in Table~\ref{tab:env_config}. Following the NIST and MITRE cybersecurity configuration~\cite{moller2023nist}, we model DDoS mitigation through four core defensive operations abstracted as logical primitives as shown in Example~\ref{ex: Logical Primitives}. We compare TAPA with several baseline methods, all of which are trained or configured in environment \textbf{E1} and then evaluated in environments \textbf{E2--E5} without prior exposure. These baselines operate with fixed action spaces consisting of one program per abstract intent defined in the example, denoted as $\mathcal{A} = \{P_{0,0}, P_{1,0}, P_{2,0}, P_{3,0}\}$. In contrast, while TAPA also starts with the same initial program set $\mathcal{A}$, it activates action adaptation for the meta-agent when abnormal network performance is detected across consecutive time windows to optimize or replace the equipped programs.

\begin{table}[h]
\centering
\small
\begin{tabular}{lccccc}
\toprule
\textbf{Env} & \textbf{Servers} & \textbf{Bots} & \textbf{Clients} & \textbf{Attack Type} & \\
\midrule
E1 & 10 & 10 & 30 & TCP Flood  \\
E2 & 10 & 10 & 30 & UDP Flood  \\
E3 & 10 & 20 & 30 & Mixed Attack \\
E4 & 10 & 10 & 80 & UDP Flood  \\
E5 & 20 & 40 & 80 & UDP Flood \\
\bottomrule
\end{tabular}
\caption{Environmental Evolution Scenarios Configuration}
\label{tab:env_config}
\end{table}

\begin{table*}[h]
\centering
\small
\setlength{\tabcolsep}{5pt}
\renewcommand{\arraystretch}{1.05}
\begin{tabular}{l *{5}{cc} c}
\toprule
\textbf{Method} 
& \multicolumn{2}{c}{\textbf{E1}} 
& \multicolumn{2}{c}{\textbf{E2}} 
& \multicolumn{2}{c}{\textbf{E3}} 
& \multicolumn{2}{c}{\textbf{E4}} 
& \multicolumn{2}{c}{\textbf{E5}} 
& \textbf{Overall Network Uptime} \\
\cmidrule(lr){2-3} \cmidrule(lr){4-5} \cmidrule(lr){6-7} \cmidrule(lr){8-9} \cmidrule(lr){10-11}
 & \textbf{Acc} & \textbf{FP} 
 & \textbf{Acc} & \textbf{FP} 
 & \textbf{Acc} & \textbf{FP} 
 & \textbf{Acc} & \textbf{FP} 
 & \textbf{Acc} & \textbf{FP} 
 &  \\
\midrule
Static Symbolic  & 100.0 & 16.7  & 100.0 & 36.7 & 100.0 & 33.3 & 70.0 & 15.0  & 15.0 & 15.0 & 48.1\\
Symbolic-Classic   & 90.0 & 10.0 & 100.0 & 30.0 & 85.0 & 50.0 & 80.0 & 45.0 & 92.5 & 87.5 & 55.7\\
Symbolic-Neural    & 100.0 & 6.7 & 100.0 & 16.7 & 100.0 & 43.3 & 60.0 & 63.8 & 97.5 & 56.3  & 70.5\\
End-to-End LLM   & 60.0 & 43.3 & 70.0 & 26.7 & 65.0 & 33.3 & 60.0 & 27.5 & 45.0  & 46.3 &  27.6  \\
TAPA (Ours)    & 100.0 & 0.0 & 100.0 & 0.0 & 100.0 & 0.0 & 100.0 & 0.0 & 100.0 & 12.5  & \textbf{77.7} \\
\bottomrule
\end{tabular}
\caption{Detection Accuracy, False Positives, and Overall Network Uptime (\%) across Evolutionary Environments.}
\label{tab:uptime_results}
\end{table*}

\paragraph{Baselines \& Evaluations}
The baseline methods include:  
\begin{itemize}
    \item \emph{Static Symbolic.} A symbolic rule-based system with no learning or adaptation.
    \item \emph{Symbolic-Classic.} A Q-learning agent operating over predefined symbolic actions.
    \item \emph{Symbolic-Neural} A Transformer-in-Transformer (TiT) based neural policy network~\cite{mao2022transformer}.
    \item \emph{End-to-End LLM} A GPT-4o-based agent that makes real-time decisions.
\end{itemize}
Evaluation is conducted using three key metrics: (1) \emph{Network uptime}: the percentage of time the victim server remains responsive; (2) \emph{Detection accuracy (Acc)}: the proportion of correctly identified bot nodes; (3) \emph{False positives (FP)}: the proportion of legitimate nodes mistakenly flagged as malicious.


\subsubsection{Effectiveness analysis of Action Adaptation}
\begin{figure}[htb]
\centering
\includegraphics[width=\columnwidth]{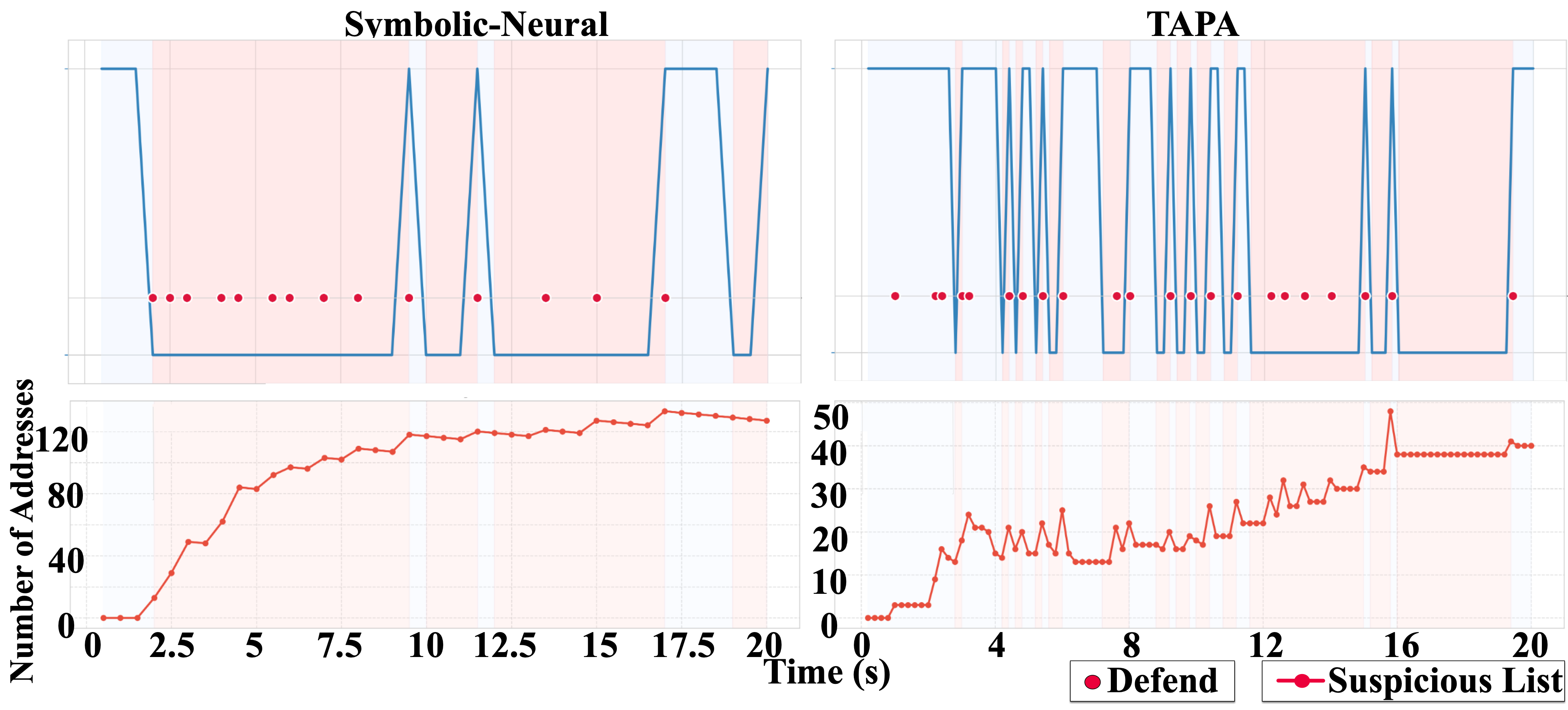}
\caption{Temporal evolution of defensive patterns under static vs. adaptive action spaces. Blue regions indicate normal state; red regions indicate attack periods. }
\label{fig:temporal evolution}
\end{figure}

Table~\ref{tab:uptime_results} validates our action adaptation strategy, demonstrating that TAPA achieves superior overall uptime without costly retraining compared to policy adaptation approaches. While pure symbolic AI and neural-symbolic methods show promising results in baseline environments (E1), they exhibit gradual performance degradation in evolved environments (E2-E5) without rule redefinition or policy retraining, highlighting the brittleness of fixed action spaces. Figure~\ref{fig:temporal evolution} further illustrates this contrast: the fixed neural-symbolic approach exhibits sparse defense activations concentrated around attack peaks, reflecting static symbolic modules' limitations under increased environmental complexity. Although policy retraining can relearn alternative strategies, such as sustained defend or continuous validate behaviors for improved network management, it lacks the timeliness and responsiveness of our action adaptation mechanism. In contrast, TAPA achieves adaptive and frequent defensive activations through dynamic symbolic program synthesis, with notably superior suspicious list management that maintains dynamic regulation rather than monotonic growth.

Overall, our \tool framework introduces LLM generalization capability to facilitate action adaptation under evolving environments without over-reliance on perfectly predefined symbolic modules. The adaptive mechanism consistently maintains 
higher performance across dynamic cybersecurity environments through context-aware program synthesis and refinement, while the meta-policy enables rapid execution. Its balanced, neuro-symbolic paradigm addresses the inherent limitations of static action spaces, constrained generalization capacity, and inefficient decision-making processes, providing a principled and scalable solution for training-free autonomous adaptation in safety-critical dynamic environments.

\begin{figure*}[h]
   \centering
   \begin{subfigure}[b]{0.49\textwidth}
       \centering
       \includegraphics[width=\textwidth]{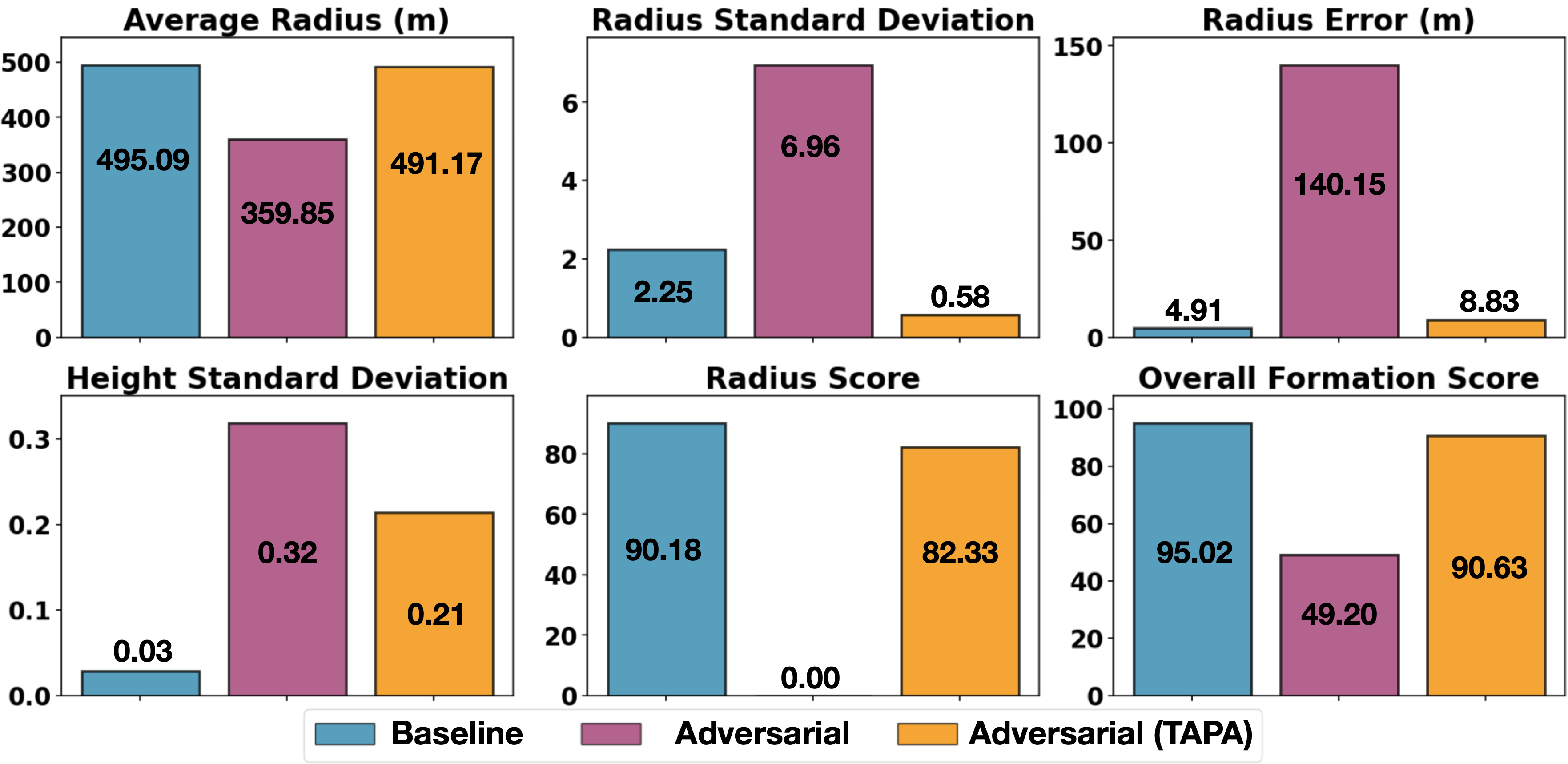}
       \caption{Adversarial scenario with malicious UAV attacks}
       \label{fig:adversarial_performance}
   \end{subfigure}
   \hfill
   \begin{subfigure}[b]{0.48\textwidth}
       \centering
       \includegraphics[width=\textwidth]{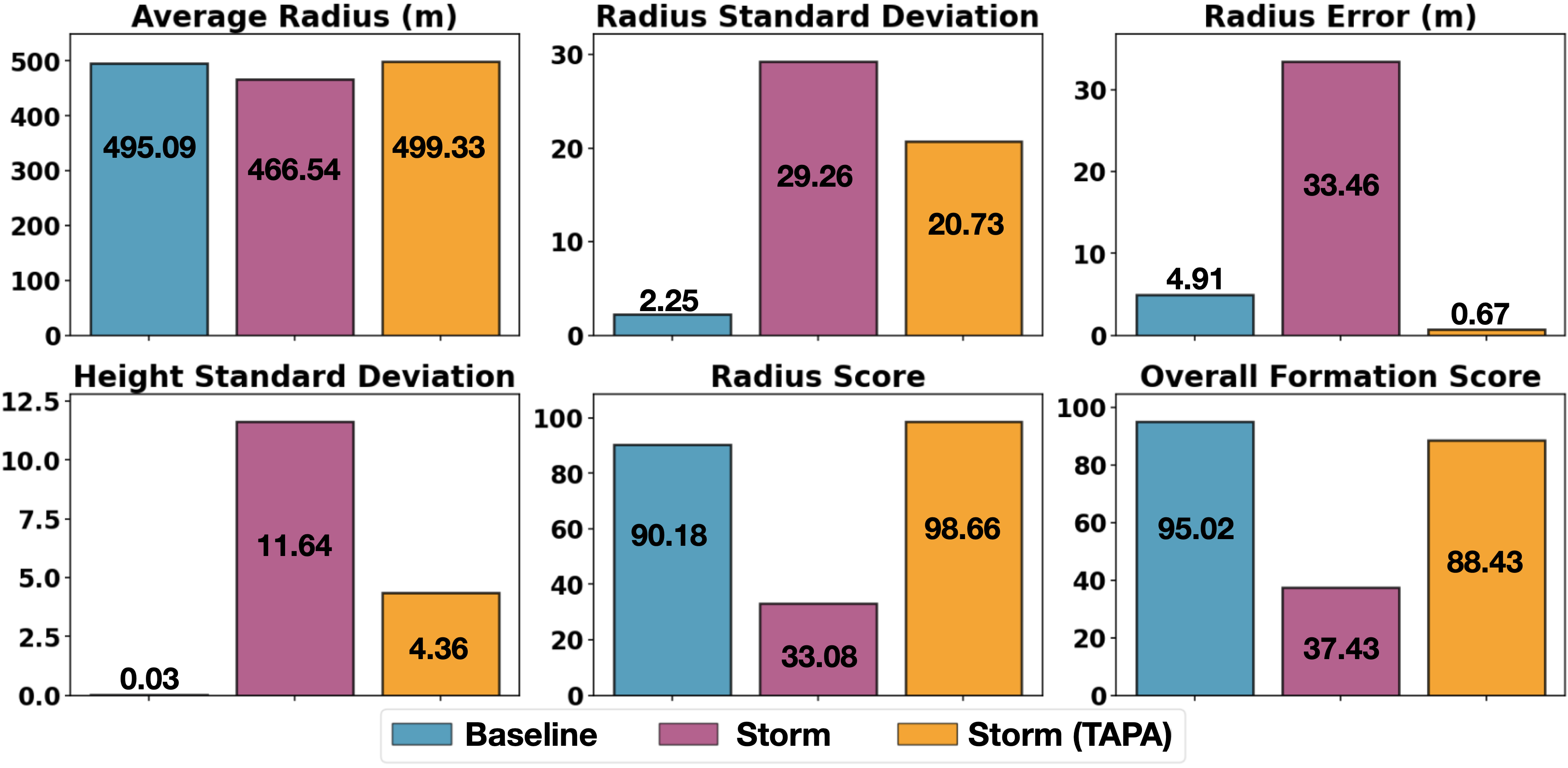}
       \caption{Storm weather scenario with environmental disturbances}
       \label{fig:storm_performance}
   \end{subfigure}
   \caption{Formation control performance across changing scenarios. TAPA consistently recovers performance in both (a) adversarial environments with malicious interference and (b) severe weather conditions, maintaining near-baseline formation quality.}
   \label{fig:uav_performance}
\end{figure*}

\subsection{Formation Control of Swarm Intelligence}
Formation control of swarm intelligence enables multiple autonomous agents (e.g., aircraft or robots) to maintain targeted formations through well-defined control algorithms. However, even carefully designed algorithms face significant limitations in dynamic environments with changing conditions, environmental disturbances, and adversarial threats. TAPA addresses these challenges by enabling real-time program synthesis for swarm coordination, dynamically adapting control parameters and logic. We evaluate TAPA against baseline methods across diverse scenarios, demonstrating its effectiveness in maintaining formation coherence under varying operational conditions.

\subsubsection{Experimental Setup}
We evaluate TAPA in multi-aircraft formation control, where 10 aircraft maintain circular formation while adapting to environmental changes and adversarial threats. The baseline strategies follow Reynolds' flocking rules~\cite{reynolds1987flocks, braga2017collision, olfati2004consensus} and defensive strategies as implemented in~\cite{hu2025enhancingrobustnessllmdrivenmultiagent}. Performance is evaluated across scenarios in Table~\ref{tab:uav_scenarios} using formation radius and height metrics. We model formation control through two logical primitives: $L_1$ (Formation Control) for adjusting coordination parameters like cohesion distance and behavioral weights, and $L_2$ (Defend) for implementing defensive programs including outlier detection and noise injection for adversarial resilience. Each aircraft uses classical flocking behaviors (separation, cohesion, alignment, goal-seeking) through distributed decision-making within communication ranges, while TAPA enables real-time adjustment of formation variables including behavioral weights, distance thresholds, and defensive programs. When overall formation score based on radius and height continuously below threshold, TAPA's LLM moderator analyzes environmental context and synthesizes appropriate parameter or program adaptations guided by expert knowledge and provenance chain experiences.

\begin{table}[h]
\centering
\small
\begin{tabular}{l p{1cm} p{5.5cm}}
\toprule
\textbf{Env} & \textbf{Aircraft} & \textbf{Scenario description} \\
\midrule
E1  & 10 & Stable weather, No any disturbance \\
E2  & 10 & Severe weather with wind and rain affecting flight dynamics \\
E3  & 10 & Malicious aircraft propagating infection to disrupt neighbor decisions \\
\bottomrule
\end{tabular}
\caption{Environmental Scenarios Configuration}
\label{tab:uav_scenarios}
\end{table}

\subsubsection{Formation Adaptation Analysis}
\begin{figure}[htb]
\centering
\includegraphics[width=\columnwidth]{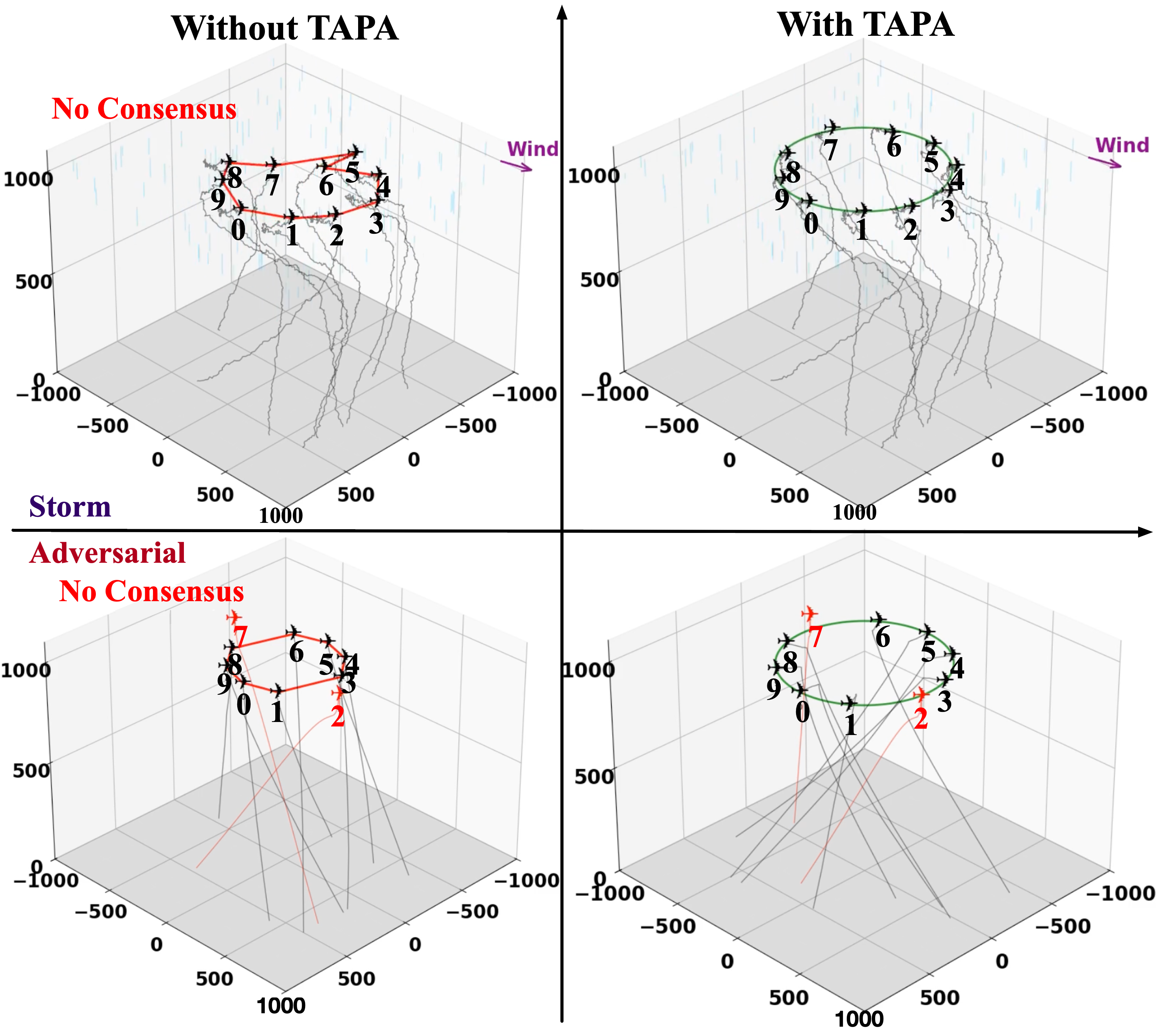}
\caption{Consensus visualization of aircraft formation.}
\label{fig:formation_adaptation}
\end{figure}

Figure~\ref{fig:formation_adaptation} and Appendix C visualizes consensus achievement across scenarios, comparing baseline formation control, scenarios without TAPA, and TAPA-enabled adaptation through 3D trajectory visualization. Figure~\ref{fig:uav_performance} provides quantitative analysis, illustrating TAPA's consistent performance recovery across adversarial attacks and storm weather conditions. The results demonstrate that while baselines fail to maintain formation consensus under environmental changes, TAPA successfully triggers multiple action adaptations in dynamic environments, continuously adjusting programs and parameters to preserve aircraft consensus. TAPA's ability to maintain near-baseline performance while adapting to diverse operational challenges validates symbolic program synthesis as an effective approach for safety-critical swarm intelligence applications requiring effectiveness, interpretability, and adaptiveness.

\subsection{Ablation Study}
To validate the design rationale of our framework components, we conduct ablation studies on formation control under storm conditions, where the system dynamically adjusts aircraft's programs and parameters across multiple adaptation rounds.  Table \ref{tab:ablation_study} reveals the complementary roles of Expert Knowledge (EK) and PC in adaptation effectiveness. Without EK, the system exhibits trial-and-error learning patterns, starting poorly but gradually improving through iterative discovery. Conversely, without PC, the system achieves reasonable initial performance due to expert-guided program generation but lacks systematic experience accumulation, causing performance to degrade over time as it cannot adapt upon previous experiences. When relying purely on LLM reasoning capabilities without either component, the system fundamentally fails to achieve meaningful improvement, as evidenced by near-zero performance throughout all rounds. These results demonstrates our framework achieved sustained improvement by combining domain expertise for reliable initialization with systematic experience learning mechanisms for continuous adaptation.
\begin{table}[htbp]
\centering
\small
\begin{tabular}{lccc}
\toprule
\textbf{Configuration} & \textbf{Round 1} & \textbf{Round 2} & \textbf{Round 3} \\
\midrule
\textbf{TAPA-Full} & 72.3 (+72.3) & 81.7 (+9.4) & 86.2 (+4.5) \\
\textbf{w/o PC} & 65.1 (+65.1)  & 68.9 (+3.8) & 62.4 (-6.5) \\
\textbf{w/o EK} & 38.6 (+38.6) & 54.2 (+15.6) & 62.7 (+8.5) \\
\textbf{w/o Both} & 4.2 (+4.2) & 6.8 (+2.6) & 0.0 (-6.8) \\
\bottomrule
\end{tabular}
\caption{Multi-Round Ablation on Formation Adaptation}
\label{tab:ablation_study}
\end{table}



\section{Conclusion}
We presented TAPA, a novel framework for training-free adaptation of programmatic agents through LLM-guided action space moderation. By positioning LLMs as intelligent moderators rather than direct decision-makers, TAPA achieves superior adaptability without costly retraining while preserving interpretability and efficiency. Our framework decouples strategic intent from execution through logical primitive abstraction, defining execution as interpretable programmatic actions to enable dynamic symbolic program synthesis in response to environmental changes. Experimental validation across cyber defense and swarm intelligence demonstrates TAPA's effectiveness in real-world scenarios. This paradigm opens new avenues for developing adaptive autonomous systems that evolve with their environments while maintaining reliability for symbolic modules.

\section*{Acknowledgments}
This work is partially funded by the European Union (under grant agreement ID 101212818). Views and opinions expressed are however those of the author(s) only and do not necessarily reflect those of the European Union or European Health and Digital Executive Agency (HADEA). Neither the European Union nor the granting authority can be held responsible for them. 
This work is partially supported by Innovate UK through AI-PASSPORT under Grant 10126404. 
Yi's contribution is partially supported through the Royal Society international exchanges programme and in part by the Engineering and Physical Sciences Research Council, through funding from RAi UK [EP/Y009800/1]. 

\bibliography{aaai2026}

@article{stodola2025dynamic,
  title={Dynamic reconnaissance operations with UAV swarms: adapting to environmental changes},
  author={Stodola, Petr and Nohel, Jan and Hor{\'a}k, Luk{\'a}{\v{s}}},
  journal={Scientific Reports},
  volume={15},
  number={1},
  pages={15092},
  year={2025},
  publisher={Nature Publishing Group UK London}
}

@inproceedings{dehghantanha2023autonomous,
  title={Autonomous cybersecurity: Evolving challenges, emerging opportunities, and future research trajectories},
  author={Dehghantanha, Ali and Yazdinejad, Abbas and Parizi, Reza M},
  booktitle={Proceedings of the Workshop on Autonomous Cybersecurity},
  pages={1--10},
  year={2023}
}

@INPROCEEDINGS{castro2025large,
  author={Castro, Sebastián R. and Campbell, Roberto and Lau, Nancy and Villalobos, Octavio and Duan, Jiaqi and Cardenas, Alvaro A.},
  booktitle={CAI}, 
  title={Large Language Models are Autonomous Cyber Defenders}, 
  year={2025},
  volume={},
  number={},
  pages={1125-1132},
  doi={10.1109/CAI64502.2025.00195}}

@article{lohn2023autonomous,
  title={Autonomous Cyber Defense},
  author={Lohn, Andrew and Knack, Anna and Burke, Ant and Jackson, Krystal},
  journal={A roadmap from lab to ops. Online. Centre for Emerging Technology and Security (CETaS) at The Alan Turing Institute},
  year={2023}
}

@article{kiran2021deep,
  title={Deep reinforcement learning for autonomous driving: A survey},
  author={Kiran, B Ravi and Sobh, Ibrahim and Talpaert, Victor and Mannion, Patrick and Al Sallab, Ahmad A and Yogamani, Senthil and P{\'e}rez, Patrick},
  journal={IEEE transactions on intelligent transportation systems},
  volume={23},
  number={6},
  pages={4909--4926},
  year={2021},
  publisher={IEEE}
}

@inproceedings{lyu2019sdrl,
  title={SDRL: interpretable and data-efficient deep reinforcement learning leveraging symbolic planning},
  author={Lyu, Daoming and Yang, Fangkai and Liu, Bo and Gustafson, Steven},
  booktitle={AAAI},
  volume={33},
  number={01},
  pages={2970--2977},
  year={2019}
}

@article{marton2024sympol,
  title={SYMPOL: Symbolic Tree-Based On-Policy Reinforcement Learning},
  author={Marton, Sascha and Grams, Tim and Vogt, Florian and L{\"u}dtke, Stefan and Bartelt, Christian and Stuckenschmidt, Heiner},
  journal={arXiv preprint arXiv:2408.08761},
  year={2024}
}

@article{bhuyan2024neuro,
  title={Neuro-symbolic artificial intelligence: a survey},
  author={Bhuyan, Bikram Pratim and Ramdane-Cherif, Amar and Tomar, Ravi and Singh, TP},
  journal={Neural Computing and Applications},
  volume={36},
  number={21},
  pages={12809--12844},
  year={2024},
  publisher={Springer}
}

@article{li2024logicity,
  title={LogiCity: Advancing neuro-symbolic ai with abstract urban simulation},
  author={Li, Bowen and Li, Zhaoyu and Du, Qiwei and Luo, Jinqi and Wang, Wenshan and Xie, Yaqi and Stepputtis, Simon and Wang, Chen and Sycara, Katia and Ravikumar, Pradeep and others},
  journal={Advances in Neural Information Processing Systems},
  volume={37},
  pages={69840--69864},
  year={2024}
}

@misc{hu2025trustorientedadaptiveguardrailslarge,
      title={Trust-Oriented Adaptive Guardrails for Large Language Models}, 
      author={Jinwei Hu and Yi Dong and Xiaowei Huang},
      year={2025},
      eprint={2408.08959},
      archivePrefix={arXiv},
      primaryClass={cs.AI},
      url={https://arxiv.org/abs/2408.08959}, 
}

@article{hadi2023survey,
  title={A survey on large language models: Applications, challenges, limitations, and practical usage},
  author={Hadi, Muhammad Usman and Qureshi, Rizwan and Shah, Abbas and Irfan, Muhammad and Zafar, Anas and Shaikh, Muhammad Bilal and Akhtar, Naveed and Wu, Jia and Mirjalili, Seyedali and others},
  journal={Authorea Preprints},
  year={2023},
  publisher={Authorea}
}

@inproceedings{guo2024large,
  title={Large language model based multi-agents: a survey of progress and challenges},
  author={Guo, Taicheng and Chen, Xiuying and Wang, Yaqi and Chang, Ruidi and Pei, Shichao and Chawla, Nitesh V and Wiest, Olaf and Zhang, Xiangliang},
  booktitle={IJCAI},
  pages={8048--8057},
  year={2024}
}

@article{hager2024evaluation,
  title={Evaluation and mitigation of the limitations of large language models in clinical decision-making},
  author={Hager, Paul and Jungmann, Friederike and Holland, Robbie and Bhagat, Kunal and Hubrecht, Inga and Knauer, Manuel and Vielhauer, Jakob and Makowski, Marcus and Braren, Rickmer and Kaissis, Georgios and others},
  journal={Nature medicine},
  volume={30},
  number={9},
  pages={2613--2622},
  year={2024},
  publisher={Nature Publishing Group US New York}
}

@inproceedings{holler2020hddl,
  title={HDDL: An extension to PDDL for expressing hierarchical planning problems},
  author={H{\"o}ller, Daniel and Behnke, Gregor and Bercher, Pascal and Biundo, Susanne and Fiorino, Humbert and Pellier, Damien and Alford, Ron},
  booktitle={Proceedings of the AAAI conference on artificial intelligence},
  volume={34},
  number={06},
  pages={9883--9891},
  year={2020}
}

@inproceedings{aineto2018learning,
  title={Learning STRIPS action models with classical planning},
  author={Aineto, Diego and Jim{\'e}nez, Sergio and Onaindia, Eva},
  booktitle={Proceedings of the International Conference on Automated Planning and Scheduling},
  volume={28},
  pages={399--407},
  year={2018}
}

@inproceedings{hogg2009learning,
  title={Learning Hierarchical Task Networks for Nondeterministic Planning Domains.},
  author={Hogg, Chad and Kuter, Ugur and Munoz-Avila, H{\'e}ctor},
  booktitle={IJCAI},
  pages={1708--1714},
  year={2009}
}

@inproceedings{kelly2008offline,
  title={Offline planning with hierarchical task networks in video games},
  author={Kelly, John-Paul and Botea, Adi and Koenig, Sven},
  booktitle={Proceedings of the AAAI Conference on Artificial Intelligence and Interactive Digital Entertainment},
  volume={4},
  number={1},
  pages={60--65},
  year={2008}
}

@article{bylander1994computational,
  title={The computational complexity of propositional STRIPS planning},
  author={Bylander, Tom},
  journal={Artificial Intelligence},
  volume={69},
  number={1-2},
  pages={165--204},
  year={1994},
  publisher={Elsevier}
}

@inproceedings{yang2018peorl,
  title={PEORL: integrating symbolic planning and hierarchical reinforcement learning for robust decision-making},
  author={Yang, Fangkai and Lyu, Daoming and Liu, Bo and Gustafson, Steven},
  booktitle={IJCAI},
  pages={4860--4866},
  year={2018}
}

@inproceedings{shindo2025blendrl,
  title     = {BlendRL: A Framework for Merging Symbolic and Neural Policy Learning},
  author    = {Shindo, Hikaru and Delfosse, Quentin and Dhami, Devendra Singh and Kersting, Kristian},
  booktitle = {Proceedings of the 13th International Conference on Learning Representations},
  year      = {2025},
}

@article{hakim2025ansr,
  title={ANSR-DT: An Adaptive Neuro-Symbolic Learning and Reasoning Framework for Digital Twins},
  author={Hakim, Safayat Bin and Adil, Muhammad and Velasquez, Alvaro and Song, Houbing Herbert},
  journal={arXiv preprint arXiv:2501.08561},
  year={2025}
}

@ARTICLE{11077439,
  author={Hu, Jinwei and Tang, Zezhi and Jin, Xin and Zhang, Benyuan and Dong, Yi and Huang, Xiaowei},
  journal={IEEE Transactions on Industrial Cyber-Physical Systems}, 
  title={Hierarchical Testing With Rabbit Optimization for Industrial Cyber-Physical Systems}, 
  year={2025},
  volume={3},
  number={},
  pages={472-484},
  doi={10.1109/TICPS.2025.3586988}}

@inproceedings{
yao2023react,
title={ReAct: Synergizing Reasoning and Acting in Language Models},
author={Shunyu Yao and Jeffrey Zhao and Dian Yu and Nan Du and Izhak Shafran and Karthik R Narasimhan and Yuan Cao},
booktitle={The Eleventh International Conference on Learning Representations},
year={2023},
}

@article{liang2025llm,
  title={LLM-Powered AI Agent Systems and Their Applications in Industry},
  author={Liang, Guannan and Tong, Qianqian},
  journal={arXiv preprint arXiv:2505.16120},
  year={2025}
}

@InProceedings{pmlr-v235-dong24c,
  title = 	 {Position: Building Guardrails for Large Language Models Requires Systematic Design},
  author =       {Dong, Yi and Mu, Ronghui and Jin, Gaojie and Qi, Yi and Hu, Jinwei and Zhao, Xingyu and Meng, Jie and Ruan, Wenjie and Huang, Xiaowei},
  booktitle = 	 {Proceedings of the 41st International Conference on Machine Learning},
  pages = 	 {11375--11394},
  year = 	 {2024},
  volume = 	 {235},
  series = 	 {Proceedings of Machine Learning Research},
  month = 	 {21--27 Jul},
  publisher =    {PMLR},
  url = 	 {https://proceedings.mlr.press/v235/dong24c.html},
}

@article{dong2024safeguardinglargelanguagemodels,
  title={Safeguarding large language models: A survey},
  author={Dong, Yi and Mu, Ronghui and Zhang, Yanghao and Sun, Siqi and Zhang, Tianle and Wu, Changshun and Jin, Gaojie and Qi, Yi and Hu, Jinwei and Meng, Jie and others},
  journal={Artificial Intelligence Review},
  volume={58},
  number={12},
  pages={382},
  year={2025},
  publisher={Springer}
}

@article{roziere2023code,
  title={Code llama: Open foundation models for code},
  author={Roziere, Baptiste and Gehring, Jonas and Gloeckle, Fabian and Sootla, Sten and Gat, Itai and Tan, Xiaoqing Ellen and Adi, Yossi and Liu, Jingyu and Sauvestre, Romain and Remez, Tal and others},
  journal={LLM4Code},
  year={2024}
}

@inproceedings{el2024using,
  title={Using github copilot for test generation in python: An empirical study},
  author={El Haji, Khalid and Brandt, Carolin and Zaidman, Andy},
  booktitle={AST 2024},
  pages={45--55},
  year={2024}
}

@inproceedings{parvez-etal-2021-retrieval-augmented,
    title = "Retrieval Augmented Code Generation and Summarization",
    author = "Parvez, Md Rizwan  and
      Ahmad, Wasi  and
      Chakraborty, Saikat  and
      Ray, Baishakhi  and
      Chang, Kai-Wei",
    booktitle = "EMNLP Findings",
    month = nov,
    year = "2021",
    url = "https://aclanthology.org/2021.findings-emnlp.232/",
    doi = "10.18653/v1/2021.findings-emnlp.232",
    pages = "2719--2734",
}

@inproceedings{
zhou2023docprompting,
title={DocPrompting: Generating Code by Retrieving the Docs},
author={Shuyan Zhou and Uri Alon and Frank F. Xu and Zhengbao Jiang and Graham Neubig},
booktitle={ICLR},
year={2023},
url={https://openreview.net/forum?id=ZTCxT2t2Ru}
}

@article{guo2024redcode,
  title={Redcode: Risky code execution and generation benchmark for code agents},
  author={Guo, Chengquan and Liu, Xun and Xie, Chulin and Zhou, Andy and Zeng, Yi and Lin, Zinan and Song, Dawn and Li, Bo},
  journal={Advances in Neural Information Processing Systems},
  volume={37},
  pages={106190--106236},
  year={2024}
}

@inproceedings{zan2023large,
  title={Large Language Models Meet NL2Code: A Survey},
  author={Zan, Daoguang and Chen, Bei and Zhang, Fengji and Lu, Dianjie and Wu, Bingchao and Guan, Bei and Yongji, Wang and Lou, Jian-Guang},
  booktitle={ACL},
  pages={7443--7464},
  year={2023}
}

@article{soroush2024large,
  title={Large language models are poor medical coders—benchmarking of medical code querying},
  author={Soroush, Ali and Glicksberg, Benjamin S and Zimlichman, Eyal and Barash, Yiftach and Freeman, Robert and Charney, Alexander W and Nadkarni, Girish N and Klang, Eyal},
  journal={NEJM AI},
  volume={1},
  number={5},
  pages={AIdbp2300040},
  year={2024},
  publisher={Massachusetts Medical Society}
}

@article{hou2025enhancing,
  title={Enhancing medical coding efficiency through domain-specific fine-tuned large language models},
  author={Hou, Zhen and Liu, Hao and Bian, Jiang and He, Xing and Zhuang, Yan},
  journal={npj Health Systems},
  volume={2},
  number={1},
  pages={14},
  year={2025},
  publisher={Nature Publishing Group UK London}
}

@inproceedings{yin2020ns3,
  title={ns3-ai: Fostering artificial intelligence algorithms for networking research},
  author={Yin, Hao and Liu, Pengyu and Liu, Keshu and Cao, Liu and Zhang, Lytianyang and Gao, Yayu and Hei, Xiaojun},
  booktitle={Proceedings of the 2020 Workshop on ns-3},
  pages={57--64},
  year={2020}
}

@article{henderson2008network,
  title={Network simulations with the ns-3 simulator},
  author={Henderson, Thomas R and Lacage, Mathieu and Riley, George F and Dowell, Craig and Kopena, Joseph},
  journal={SIGCOMM demonstration},
  volume={14},
  number={14},
  pages={527},
  year={2008}
}

@article{sharafaldin2018toward,
  title={Toward generating a new intrusion detection dataset and intrusion traffic characterization.},
  author={Sharafaldin, Iman and Lashkari, Arash Habibi and Ghorbani, Ali A and others},
  journal={International Conference on Information Systems Security and Privacy},
  year={2018}
}

@article{hu2025enhancingrobustnessllmdrivenmultiagent,
title = {Enhancing robustness of LLM-driven multi-agent systems through randomized smoothing},
journal = {Chinese Journal of Aeronautics},
pages = {103779},
year = {2025},
issn = {1000-9361},
doi = {https://doi.org/10.1016/j.cja.2025.103779},
url = {https://www.sciencedirect.com/science/article/pii/S1000936125003851},
author = {Jinwei HU and Yi DONG and Zhengtao DING and Xiaowei HUANG}
}

@inproceedings{reynolds1987flocks,
  title={Flocks, herds and schools: A distributed behavioral model},
  author={Reynolds, Craig W},
  booktitle={Annual conference on Computer graphics and interactive techniques},
  pages={25--34},
  year={1987}
}

@inproceedings{braga2017collision,
  title={Collision avoidance based on Reynolds rules: A case study using quadrotors},
  author={Braga, Rafael G and Da Silva, Roberto C and Ramos, Alexandre CB and Mora-Camino, Felix},
  booktitle={Information Technology-New Generations: 14th International Conference on Information Technology},
  pages={773--780},
  year={2017},
  organization={Springer}
}

@article{olfati2004consensus,
  title={Consensus problems in networks of agents with switching topology and time-delays},
  author={Olfati-Saber, Reza and Murray, Richard M},
  journal={IEEE Transactions on automatic control},
  volume={49},
  number={9},
  pages={1520--1533},
  year={2004},
  publisher={IEEE}
}

@inproceedings{duan2025enhancing,
  title={Enhancing multi-agent consensus through third-party llm integration: Analyzing uncertainty and mitigating hallucinations in large language models},
  author={Duan, Zhihua and Wang, Jialin},
  booktitle={International Conference on Advanced Algorithms and Control Engineering },
  pages={2222--2227},
  year={2025},
  organization={IEEE}
}

@article{verma2019imitation,
  title={Imitation-projected programmatic reinforcement learning},
  author={Verma, Abhinav and Le, Hoang and Yue, Yisong and Chaudhuri, Swarat},
  journal={Advances in Neural Information Processing Systems},
  volume={32},
  year={2019}
}

@inproceedings{carnevali2024neuro,
  title={Neuro-Symbolic Artificial Intelligence for Safety Engineering},
  author={Carnevali, Laura and Lippi, Marco},
  booktitle={International Conference on Computer Safety, Reliability, and Security},
  pages={438--445},
  year={2024},
  organization={Springer}
}

@article{kuchta2018shadow,
  title={Shadow symbolic execution for testing software patches},
  author={Kuchta, Tomasz and Palikareva, Hristina and Cadar, Cristian},
  journal={TOSEM},
  volume={27},
  number={3},
  pages={1--32},
  year={2018},
  publisher={ACM New York, NY, USA}
}

@misc{microsoft2023shadow,
  title={Shadow Testing - Engineering Fundamentals Playbook},
  author={{Microsoft}},
  year={2023},
  howpublished={\url{https://microsoft.github.io/code-with-engineering-playbook/automated-testing/shadow-testing/}},
}

@article{achiam2023gpt,
  title={Gpt-4 technical report},
  author={Achiam, Josh and Adler, Steven and Agarwal, Sandhini and Ahmad, Lama and Akkaya, Ilge and Aleman, Florencia Leoni and Almeida, Diogo and Altenschmidt, Janko and Altman, Sam and Anadkat, Shyamal and others},
  journal={arXiv preprint arXiv:2303.08774},
  year={2023}
}

@misc{claude3,
  author = {Anthropic},
  title = {Claude 3 Technical Report},
  year = {2024},
  url = {https://www.anthropic.com/news/claude-3-family},
}

@article{jiang2024survey, author = {Jiang, Juyong and Wang, Fan and Shen, Jiasi and Kim, Sungju and Kim, Sunghun}, title = {A Survey on Large Language Models for Code Generation}, year = {2025}, publisher = {Association for Computing Machinery}, address = {New York, NY, USA}, issn = {1049-331X}, url = {https://doi.org/10.1145/3747588}, doi = {10.1145/3747588}, note = {Just Accepted}, journal = {TOSEM}, month = jul }

@misc{hu2025stopreducingresponsibilityllmpowered,
      title={Stop Reducing Responsibility in LLM-Powered Multi-Agent Systems to Local Alignment}, 
      author={Jinwei Hu and Yi Dong and Shuang Ao and Zhuoyun Li and Boxuan Wang and Lokesh Singh and Guangliang Cheng and Sarvapali D. Ramchurn and Xiaowei Huang},
      year={2025},
      eprint={2510.14008},
      archivePrefix={arXiv},
      primaryClass={cs.MA},
      url={https://arxiv.org/abs/2510.14008}, 
}

@inproceedings{lelis2024programmatic,
  title     = {Generating Programmatic Solutions: Algorithms and Applications of Programmatic Reinforcement Learning and Code Generation},
  author    = {Levi Lelis and Xinyun Chen and Shao‑Hua Sun},
  booktitle = {NeurIPS 2024 Tutorial},
  year      = {2024},
  address   = {East Exhibition Hall A, NeurIPS},
}

@incollection{moller2023nist,
  title={NIST cybersecurity framework and MITRE cybersecurity criteria},
  author={M{\"o}ller, Dietmar PF},
  booktitle={Guide to Cybersecurity in Digital Transformation: Trends, Methods, Technologies, Applications and Best Practices},
  pages={231--271},
  year={2023},
  publisher={Springer}
}

@article{mao2022transformer,
  title={Transformer in Transformer as Backbone for Deep Reinforcement Learning},
  author={Mao, Hangyu and Zhao, Rui and Chen, Hao and Hao, Jianye and Chen, Yiqun and Li, Dong and Zhang, Junge and Xiao, Zhen},
  journal={arXiv preprint arXiv:2212.14538},
  year={2022}
}

@incollection{strom2018mitre,
  title={Mitre att\&ck: Design and philosophy},
  author={Strom, Blake E and Applebaum, Andy and Miller, Doug P and Nickels, Kathryn C and Pennington, Adam G and Thomas, Cody B},
  booktitle={Technical report},
  year={2018},
  publisher={The MITRE Corporation}
}

@article{zargar2013survey,
  title={A survey of defense mechanisms against distributed denial of service (DDoS) flooding attacks},
  author={Zargar, Saman Taghavi and Joshi, James and Tipper, David},
  journal={IEEE communications surveys \& tutorials},
  volume={15},
  number={4},
  pages={2046--2069},
  year={2013},
  publisher={IEEE}
}
\appendix
\section{Implementation Details} \label{app:exec_details}
\subsection{Autonomous cyber defense}
We implement three DDoS attack patterns in the NS3 simulator to evaluate the effectiveness of action adaptation. DDoS attacks overwhelm target servers by coordinating multiple compromised nodes to generate malicious traffic that exhausts network resources or computational capacity \cite{zargar2013survey}. Traditional automated network defense systems face significant challenges when confronting these attacks due to scalability constraints, parameter uncertainty, and evolving attack patterns. Static defense mechanisms struggle with dynamic scaling requirements as attack volumes fluctuate, face difficulties in determining optimal detection thresholds without prior knowledge of attack characteristics, and exhibit brittleness when encountering novel attack vectors that deviate from predefined setting.

In our experimental configuration, we establish a baseline network infrastructure with 100 Mbps bandwidth capacity to simulate enterprise-grade connectivity. The simulation environment comprises 10-20 victim servers, 30-80 legitimate clients generating normal traffic patterns, and 10-40 coordinated botnet nodes. We include prevalent attack types that stress different aspects of network infrastructure. UDP flood attacks saturate target bandwidth by transmitting massive volumes of connection-less UDP packets at 50 Mbps per bot, characterized by high transmission rates and low computational overhead that evade connection state-based detection mechanisms. TCP flood attacks exploit the three-way handshake protocol to exhaust server connection resources, generating 1,000 connection attempts per second and manifesting as numerous half-open connection establishment patterns. Mixed attacks dynamically alternate between UDP and TCP characteristics, with all bots randomly activating throughout the simulation timeline, presenting greater challenges to maintaining network uptime and requiring adaptive defense strategies capable of handling multi-vector threats.

Each environment scenario E1-E5 operates with a 5-second simulation duration serving as the performance evaluation window. This temporal design balances real-time attack detection requirements with defense strategy adaptation latency, ensuring sufficient data samples for statistical significance analysis while maintaining practical deployment time constraints. TAPA's meta-agent employs a TiT neuro-symbolic architecture \cite{mao2022transformer}, trained in baseline environment E1 to learn the mapping strategy from network state space to logical primitives. The training process establishes semantic foundations for symbolic defensive actions, providing a stable decision-making framework for subsequent action space adaptation. We incorporate expert knowledge such as MITRE ATT\&CK knowledge base \cite{strom2018mitre} to guide Claude Sonnet 4, providing domain-specific guidance principles for symbolic program synthesis. When performance degradation triggers are detected within the 20-second evaluation window (network uptime $<50\%$ or active flow count exceeding three times the baseline), we transmit current environmental context, performance variations, and historical adaptation experiences from previous code modifications to Claude Sonnet 4 for symbolic program synthesis execution.

\subsection{Swarm formation control}
This experiment evaluates TAPA's runtime adaptation capabilities in coordinated multi-agent systems under environmental disturbances and adversarial interference. Swarm formation control requires multiple autonomous agents to maintain predetermined geometric configurations while navigating dynamic environments, presenting challenges in consensus achievement, collision avoidance, and robustness to external perturbations. Traditional formation control algorithms rely on fixed control parameters and static coordination strategies that become ineffective when environmental conditions deviate from design assumptions or when adversarial agents attempt to disrupt swarm cohesion.

In our experimental setup, we simulate a 10-aircraft formation maintaining circular formation in a 3D airspace environment. The baseline formation control implements classical distributed algorithms including separation, cohesion, and alignment behaviors with realistic communication constraints. We design three progressive scenarios: E1 represents stable weather conditions, E2 introduces storm patterns with wind and precipitation affecting flight dynamics, and E3 incorporates malicious aircraft that inject false information to disrupt neighbor decision-making processes. 

Each scenario operates for 500-frame evaluation windows to assess formation stability metrics. Formation quality is quantified through a composite scoring function to evaluate consensus achievement:
\begin{equation}
S_{overall} = \frac{S_{radius} + S_{height}}{2}
\end{equation}
where $S_{radius} = \max(0, 100 - \beta \cdot E_{radius})$ represents the radius component score, $S_{height} = \max(0, 100 - \gamma \cdot \sigma_{height})$ denotes the height component score, $E_{radius}$ is the radius error from target formation geometry, $\sigma_{height}$ is the altitude standard deviation across agents, and $\beta$, $\gamma$ are penalty coefficients weighting geometric precision and vertical coordination respectively.

TAPA's meta-agent operates over two logical primitives based on \cite{hu2025enhancingrobustnessllmdrivenmultiagent}: L1 (Formation Control) for adjusting coordination parameters such as cohesion distance and behavioral weights, and L2 (Defend) for implementing defensive programs including outlier detection and adversarial resilience mechanisms. The meta-agent autonomously selects appropriate logical primitives based on formation quality metrics and environmental context indicators. When formation consensus fails within 150 simulation frames, GPT-4o synthesizes adaptive control programs incorporating domain expertise in distributed control theory and multi-agent coordination principles. This enables real-time adaptation to both environmental changes and adversarial interference while maintaining interpretability and formal guarantees essential for safety-critical swarm applications.

\section{Provenance Chain Example} \label{app: Provenance chain example}
\begin{figure*}[!t]
\centering
\includegraphics[width=\textwidth]{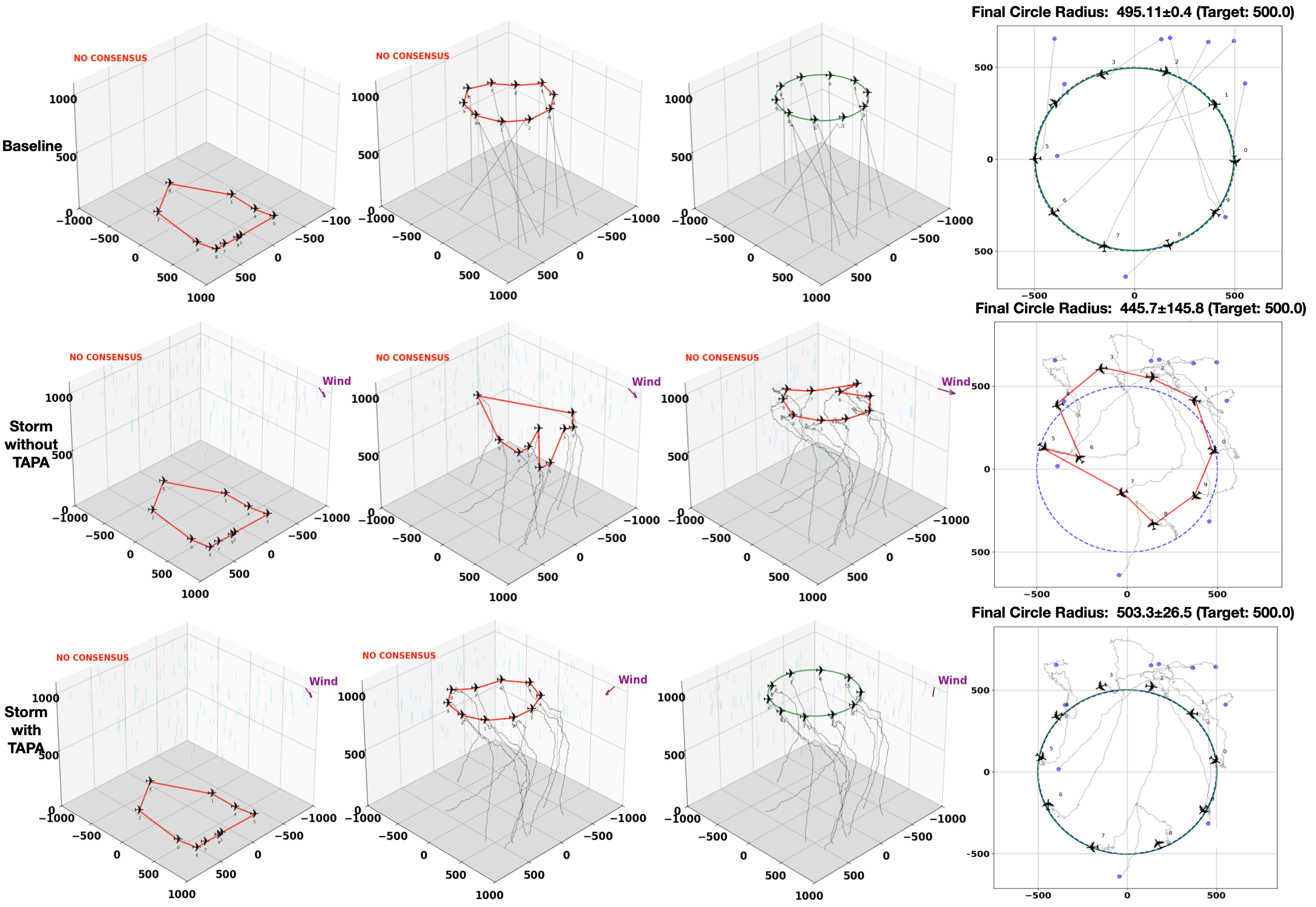}
\caption{Temporal evolution of swarm formation control under storm conditions. The figure shows three scenarios: (top) baseline formation under stable conditions, (middle) formation breakdown during storm without adaptation, and (bottom) successful formation maintenance with TAPA adaptation. Left panels show 3D trajectories over time, while right panels provide top-view final formation states. TAPA enables consistent consensus achievement despite environmental disturbances.}
\label{fig:swarm_formation_evolution}
\end{figure*}

\begin{figure*}[!ht]
\centering
\includegraphics[width=\textwidth]{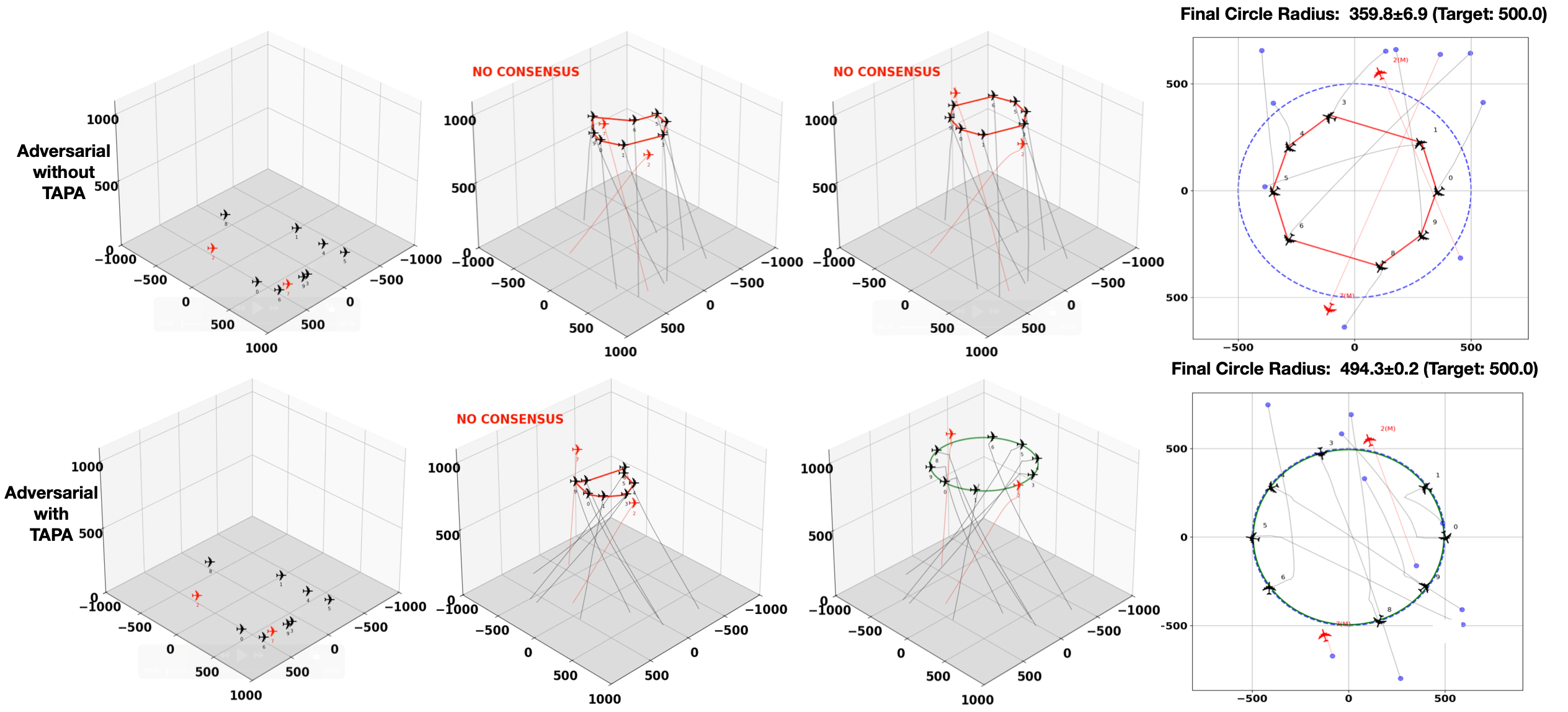}
\caption{Formation control performance under adversarial interference. The figure compares swarm behavior with and without TAPA adaptation when facing malicious aircraft attacks. Left panels show 3D trajectory evolution, while right panels provide quantitative top-view analysis of final formation quality.}
\label{fig:adversarial_formation_comparison}
\end{figure*}

\begin{example}[Provenance chain example for cyber defense]
\begin{verbatim}

{
 "timestamp": "2025-01-15 14:30:22",
 "logical_primitive": "L3: Defense",
 "environmental_context": {
   "average_delay": "50ms - 600ms",
   "traffic_volume": "1K - 10K 
   packets/sec",
   "packet loss rate": "0.0 - 0.45",
 },
"program_adaptation": {
  "original_program": "P4.0",
  "selected_program": {
    "programs": ["P4.0", "P4.1", "P4.2"]
  },
  "generated_program": null,
  "adaptation_details": {
    "parameter_changes": {
      "P4.2": "blacklist_t=0.8",
      "P4.1": "rate_limit=500", 
    }
  },
  "confidence": "95%"
},
 "validation_results": {
   "shadow_mode": "success",
   "success_rate": "95%",
   "response_time": "2.3s"
 },
 "outcome": {
   "status": "attack_mitigated",
   "performance_improvement": "67%"
 },
 "interpretable_rationale": {
   "detection_reason": "higher traffic 
   rate detected",
   "adaptation_logic": "pattern matches 
   higher DDoS scale",
   "rollback_available": true
 },
 "agent_version": "v0.0.2",
 "program_version": "ddos_v1.2",
 "rag_version": "cyber_v3.0"
}
\end{verbatim}
\end{example}

\section{Temporal Evolution Analysis in Swarm Formation Control}
Figure~\ref{fig:swarm_formation_evolution} illustrates the complete temporal evolution of swarm formation under storm conditions, demonstrating TAPA's adaptive capabilities in maintaining consensus despite environmental disturbances. The visualization presents three scenarios across multiple time snapshots: baseline operation under stable conditions, storm impact without adaptation, and storm resilience with TAPA intervention. In the baseline scenario (top row), the 10-aircraft formation maintains stable circular configuration throughout the simulation. When storm conditions are introduced without adaptive mechanisms (middle row), severe wind disturbances cause formation breakdown, with aircraft trajectories becoming increasingly chaotic and losing consensus. However, with TAPA's action adaptation enabled (bottom row), the swarm successfully recovers and maintains formation coherence despite identical storm conditions. The rightmost panels provide top-view perspectives showing the final formation quality, where TAPA-enabled agents achieve near-perfect circular formation while baseline storm scenarios result in complete dispersion.

Complementing the storm scenario analysis, Figure~\ref{fig:adversarial_formation_comparison} evaluates TAPA's performance under adversarial interference, where malicious UAVs inject false information to disrupt swarm coordination. Without adaptive mechanisms, the formation suffers significant degradation with a final radius deviation of $359.8 \pm 6.9$ units from the target 500.0, indicating substantial consensus failure. In contrast, TAPA-enabled swarms demonstrate remarkable resilience, achieving near-optimal formation with radius $494.3 \pm 0.2$, representing only 1.1\% deviation from the target. This quantitative comparison validates TAPA's ability to distinguish between environmental disturbances and adversarial attacks, synthesizing appropriate defensive strategies that preserve formation integrity. The temporal analysis validates TAPA's ability to synthesize context-aware control parameters that preserve swarm stability under diverse dynamic challenges, from natural environmental perturbations to deliberate adversarial interference.

\end{document}